\documentclass[review]{elsarticle}
\usepackage{graphics}
\usepackage{graphicx}
\usepackage{float}
\usepackage{bm}
\usepackage{rotating}
\usepackage{multirow}

\usepackage[nomarkers,notablist,nofiglist]{endfloat}

\usepackage{subfigure}

\usepackage{lineno}
\modulolinenumbers[5]
\usepackage{siunitx}         
\graphicspath{ {./Figures/} } 
\usepackage{soul}
\usepackage{miller}
\usepackage{amsmath}
\usepackage{geometry} 
\usepackage{color} 
\usepackage{outlines}
\usepackage{upgreek}
\journal{arXiv}

\begin{document}

\begin{frontmatter}
	
    \title{Depth-Resolved Characterization of Centrifugal Disk Finishing of Additively Manufactured Inconel 718}

    \author[1]{Kenneth M. Peterson}
    \author[2]{Mustafa Rifat}
    \author[2]{Edward C. DeMeter}
    \author[2]{Saurabh Basu}
    \author[1]{Darren C. Pagan\corref{mycorrespondingauthor}}
 	\ead{dcp5303@psu.edu}
    \cortext[mycorrespondingauthor]{Corresponding author}

    \address[1]{Materials Science and Engineering, The Pennsylvania State University, University Park, PA 16802}
    \address[2]{Industrial Engineering, The Pennsylvania State University, University Park, PA 16802}

\begin{abstract}
Surface characteristics are a major contributor to the in-service performance, particularly fatigue life, of additively manufactured (AM) components. Centrifugal disk finishing (CDF) is one of many rigid media, abrasive machining processes employed to smooth the surfaces and edges of AM components.  Within the general family of abrasive machining processes currently applied to AM, CDF is moderate in terms of material removal rate and the inertial forces exerted. How CDF alters the underlying microstructure of the processed surface is currently unknown. Here we employ white light profilometry and high-energy X-ray diffraction to characterize surface finish, crystallographic texture, and anisotropic distributions of residual microscale strain as a function of depth in CDF-finished Inconel 718 manufactured with laser powder bed fusion. Surfaces are finished using both unimodal and bimodal finishing media size distributions. We find that CDF will remove surface crystallographic textures (here a \{111\} fiber texture) from AM components, but generally not alter the bulk texture (here a cube texture). CDF is also found to impart significant amounts of residual microscale strain into the first 100 $\mu$m from the sample surface as evidenced by an approximately 50\% increase in diffraction peak widths at 20 $\mu$m from the surface in comparison to 120 $\mu$m.

\end{abstract}

\end{frontmatter}


\section{Introduction}

Surface integrity is a controlling feature for the mechanical performance of metallic components, particularly the fatigue life. Surfaces of additively manufactured (AM) components often have poor surface integrity in the as-built state which is naturally detrimental to the mechanical response \cite{gockel2019influence,lee2021review}. Primary machining processes are employed to subtract AM material to create functional surfaces with specific geometry and fine surface texture. For high performance components, secondary machining processes are additionally employed to subtract AM material from non-functional surfaces to smooth them. A common means to perform this secondary machining is to employ a family of processes referred to here as rigid media, abrasive machining processes. These processes are favored because they are capable of batch processing and are relatively inexpensive.

These processes employ: 1) rigid media with a designed shape, comprised of a ceramic or plastic matrix embedded with abrasive grit; 2) a machine with an enclosure that is filled with the loose media and parts that are either allowed to free-float within the enclosure, fixed to the enclosure, or attached to a mechanical arm; and 3) motion of the enclosure, motion of mechanical elements within the enclosure, or motion of the mechanical arm with attached part to generate relative motion between the media and parts with varying contact pressure and relative velocity.  
Centrifugal barrel finishing and drag finishing processes are examples of high kinetic energy processes that are commonly used to quickly strip off AM material and leave behind semi-rough surfaces. These surfaces are subsequently finished using low kinetic energy processes such as vibratory bowl finishing and barrel finishing (also known as tumbling).

\begin{figure}[h]
      \centering \includegraphics[width=0.5\textwidth]{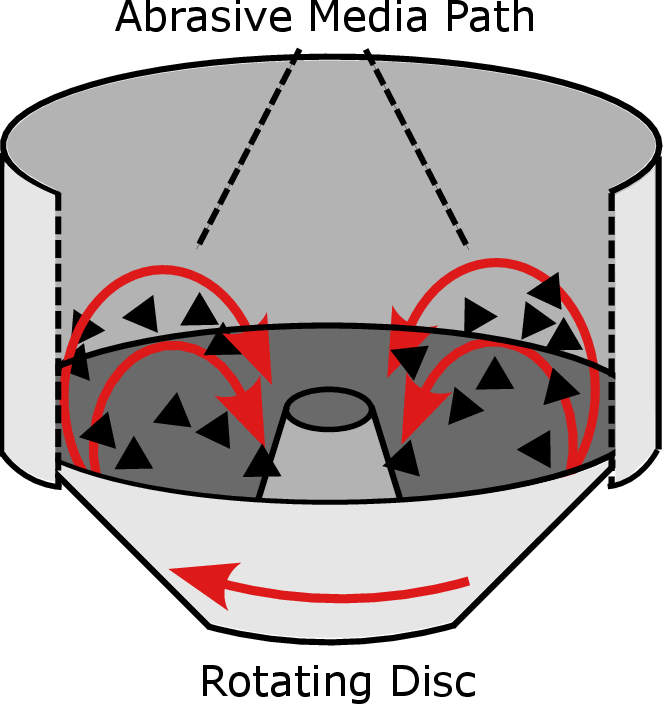}
    \caption{A schematic of the centrifugal disk finishing process in which abrasive media and components are placed into a cylindrical bowl. A rotating disc at the bottom of the assembly excites media and component motion to induce finishing.}
    \label{fig:cdf}
\end{figure}

Centrifugal disk finishing (CDF) is a mid-level, kinetic energy process that is commonly used for both semi-roughing and finishing of AM parts, with each process using a different media \cite{rifat2023surface,de2019super}. The process employs a bowl with a rotating disk at the bottom which is illustrated schematically in Fig. \ref{fig:cdf}.  Media and parts are placed unrestrained in the bowl. The rotating disk energizes the media and parts, causing both to move rapidly through the bowl. While this occurs, an emulsion flows through the mix to lubricate the media and parts and to carry away fines.   Media differ in shape, size, and composition to suit the geometry, starting surface texture, and hardness of the part. The smoothing effect in CDF is shear-based and originates from mechanical interaction between the media used and the surface. This mechanical interaction can also alter the microstructure and the stress state of the material in the vicinity of the surface of the part. In this regard, the evolution of surface integrity of a part during CDF, characterized by its surface roughness and crystallographic texture is coupled.

Although the effectiveness of rigid media, abrasive machining as a surface improvement approach is well established \cite{,kopp2022prediction,rifat2023surface,fan2023mass} even in commercial settings, fundamental details of how the process alters the material microstructure below the surface is not well explored. For instance, Nalli et al. show that barrel finishing the surface of titanium alloy Ti6Al4V uniaxial tensile test specimens improves their yield strength, ultimate tensile strength and ductility \cite{nalli2020effect}, but Denti et al. found that tumbling alone was not sufficient to improve fatigue life of laser powder bed fusion Ti6Al4V and subsequent shot-peening was required. \cite{denti2019fatigue}. Xiuhua et al. report increased surface hardness and compressive residual stresses in 18CrNiMo7-6 carburized rollers that underwent barrel finishing \cite{zhang2022experimental}. In fact, an increase in residual stress was also seen in nickel super alloy Inconel 718 that underwent barrel finishing \cite{rifat2023surface}. Together these results imply that improving surface finish alone is not sufficient to control final mechanical performance of AM parts, and the microstructure and residual stress modification also play a role. Understanding these effects can lead to optimization of process conditions beyond final surface condition and lead to superior mechanical performance.

Optimization of abrasive treatments are complicated by process mechanics which often involve severe (subsurface) plastic strains generated by stochastic interactions of the surface with the rigid abrasive media \cite{basu2016deformation,wang2018quantifying}. While the surface finish may be improved \cite{rifat2024evolution}, the microstructure and local residual strain and stress states below the surface are also altered which, in-turn, will alter the local mechanical response. Analysis of these subsurface changes have traditionally been hampered by characterization limitations and as such are generally not optimized. Destructive serial-sectioning combined with microscopy will alter residual stress state, while low-energy, laboratory X-ray diffraction cannot penetrate metallic alloys with sufficient depth. Changes in microstructure can be inferred from surface hardness testing \cite{rifat2020microstructure}, but these measurements are not a direct measure of microstructural changes. However, synchrotron high-energy X-ray diffraction provides an ideal tool for non-destructively characterizing the effects of abrasive surface treatments on AM materials for process and microstructural optimization.

Here we examine how variation of abrasive media and processing time alter surface state, along with depth-resolved subsurface microstructural and microstrain state in AM Inconel 718 (IN718) built using laser powder bed fusion. IN718 nickel superalloy is generally employed for its high-temperature strength and corrosion resistance \cite{chaturvedi1983strengthening,reed2008superalloys} and its machining \cite{rahman1997machinability,dudzinski2004review,pawade2008effect,kadam2017surface,kadam2022cutting} and finishing \cite{yao2013research,chaabani2020comparison,miranda2021comparative} response has been extensively studied. This material, along with Inconel 625, has also been a focus of intense study by the additive manufacturing community \cite{wang2017review,hosseini2019review,mostafaei2023additive}. Characterization is performed using both white light profilometry and high-energy X-ray diffraction. We find clear variation in microscale stress state with varying processing route and alteration of surface preferred crystallographic orientation due to surface material removal. In addition, we show that as surface roughness decreases, the magnitude of residual stresses increases during CDF. However, the mechanical interactions between the media and the surface that produced this effect were not strong enough to cause evolution of crystallographic textures in the material close to the surface. This implies a crystallographic texture-preservation effect, which is unexpected in shear-based finishing processes \cite{wang2018quantifying,basu2015crystallographic}. This is an important result as the surface integrity of a part, which is characterized by its roughness, residual stress, and crystallographic texture, significantly influences its ability to resist damage during fatigue. Herein, the ability to preserve beneficial crystallographic textures near the surface can provide novel avenues for enhancing a part's life, particularly during cyclic loading conditions.

This work is structured as follows. We begin by introducing the material and surface processing methods employed, followed by the characterization techniques. Results describe the final surface finish and subsurface microstructures and microstrains. Implications for the application of centrifugal disk finishing to AM processes are then discussed.

\section{Material and Characterization}

\subsection{Material and CDF Finishing}

Inconel 718 specimens were extracted from a 5 mm $\times$ 25 mm $\times$ 25 mm block(s) built using laser powder bed fusion (LPBF). The blocks were built in an EOS M280 DMLS machine using manufacturer-supplied EOS-718 powder. During the build, the laser spot size was 87.5 $\mu$m, laser power was 285 W, and the laser velocity was 960 mm/s. The layer build thickness was 40 $\mu$m and a 67.5$^\circ$ layer rotation was applied after each layer. Prior to separating the blocks from the build plate, a stress relief heat-treatment was applied within a vacuum furnace with the manufacturer settings of 1065 $^\circ$C for approximately 90 minutes (variation of temperature within was $\pm$ 12 $^\circ$C). The blocks were then removed from the build using electro discharge machining.

\begin{figure}[h]
      \centering \includegraphics[width=0.8\textwidth]{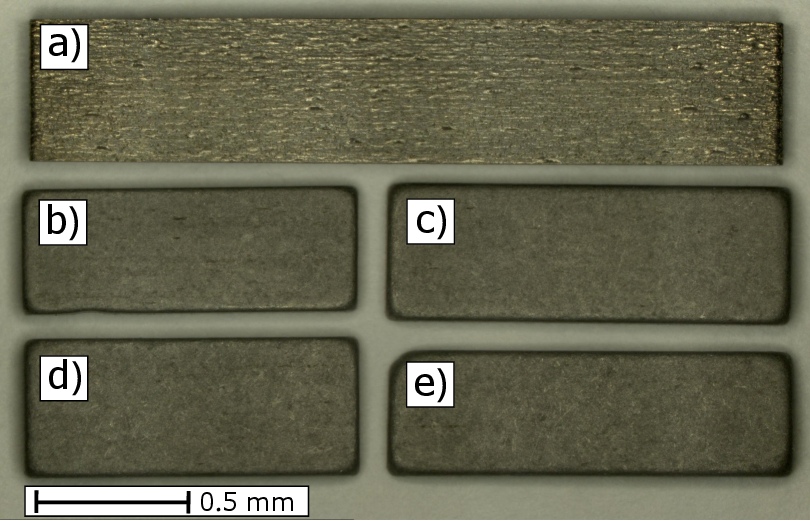}
    \caption{A photo of the 5 surface-finish specimens characterized with high-energy X-ray diffraction. a) NF (not finished), b) U1 (6$\times$6 SCT media, 75 mins), c) U2 (6$\times$6 SCT media, 135 mins), d) B1 (bimodal media, 75 mins), e) B2 (bimodal media, 135 mins).}
    \label{fig:pics}
\end{figure}

After removal, blocks were placed into a centrifugal finisher with triangular prism shaped ceramic (Al2O3) media obtained from Walther Trowal. Two media sizes were used with approximately 6 mm (WXC 6$\times$6 straight cut triangle, SCT) and 15 mm (WXC 15$\times$15 angle cut triangle, ACT) particle sizes. Four specimens were separately placed into a Walther Trowal TT45 centrifugal disc finisher  under different surface processing conditions. Finishing was carried out in a lubricated condition. The KFL compound purchased from Walther Trowal was used as a lubricant and endowed to the media at the rate 300 mL/min. The speed of revolution for finishing was 250 rpm for all the specimens. A cumulative total amount of 25 kg of media was used for both, unimodal and bimodal conditions. Two specimens were processed in only the 6 mm particles (unimodal mixture) for 75 minutes and 135 minutes respectively (U1 and U2), while another two specimens were processed in a 50:50 mixture of 6 mm and 15 mm particles (bimodal mixture) also for 75 minutes and 135 minutes respectively (B1 and B2). One specimen was reserved with no surface finish (NF). After processing 0.9 mm thick specimens were extracted from the larger blocks. In total, 5 specimens were generated and the processing parameters are summarized in Table \ref{tab:rough_table}. A photo of the 5 specimens is provided in Fig. \ref{fig:pics} where the finishing effect of the CDF process can be clearly visualized between the NF and finished specimens.

\begin{table}[h]
\footnotesize
\centering

\caption{Summary of Inconel 718 specimens produced using laser powder bed fusion  characterized in this study.}

\begin{tabular}{|m{2cm}|m{3cm}|m{1cm}|m{1.8cm}|}
\hline
\textbf{Finish Type} & \textbf{Media} & \textbf{Name} & \textbf{Processing Time (min.)}\\

\hline
No Finish & N/A & NF & N/A\\
\hline
\multirow{2}{6em}{Unimodal} & \multirow{2}{9em}{WXC $6\times6$ SCT} & U1 & 75\\
&& U2 &135 \\
\hline

\multirow{2}{6em}{Bimodal} & \multirow{2}{9em}{WXC $6\times6$ SCT + WXC $15\times15$ ACT} &  B1 & 75\\
&& B2 &135 \\
\hline

\end{tabular}

\label{tab:rough_table}
\end{table}

\subsection{Surface Roughness Characterization}
Surface profiles of the NF and centrifugal-disc-finished specimens were characterized by white light profilometry using a Zygo Nexview 3D Optical Surface Profiler. The optical setup of this white-light surface profiler comprised a 20$\times$ -0.5$\times$ magnification that provided readings over a physical scan area of 836 $\mu$m $\times$ 836 $\mu$m that were recorded over a final digital area of size 1024 pixels $\times$ 1024 pixels in each case (scan resolution of $\sim$ 0.82 $\mu$m/pixel). No digital filtering was applied to the profilometry data in post-processing or analysis.

From these scans, multiple measures of surface roughness $R$ were extracted including arithmetic average $R_a$
\begin{equation}
R_a=\frac{1}{n}\sum_{i=1}^n|z_i - \bar{z}| \quad ,
\end{equation}
root mean square $R_q$
\begin{equation}
R_q=\sqrt{\frac{1}{n}\sum_{i=1}^n (z_i- \bar{z})^2} \quad , 
\end{equation}
and maximum peak-to-valley height $R_z$
\begin{equation}
R_z=\max{(z_i)}-\min{(z_i)}
\end{equation}
where $z_i$ is the height at each pixel in the surface roughness map, $\bar{z}$ is the mean value of $z_i$, and $n$ is the number of surface height measurements. Three positions were randomly chosen on each specimen, and the average of the surface roughness measures obtained from these spots are reported. The orientations of these surface normals were always perpendicular to the build direction.

The choice of these parameters for the current study was motivated by their widespread use in industry. Among these, benchmark values for the $R_a$ parameter for various applications such as sealing, and fatigue loading are well known. This makes it easier to qualify a part for a certain application, e.g., if its $R_a$ value is smaller than the pertinent benchmark. Nonetheless, these parameters provide an incomplete picture of the surface \cite{rifat2020microstructure}. More sophisticated parameters that have a systematically calculable physical meaning, e.g., surface curvature, can be used to achieve a more comprehensive description of the condition of the surface \cite{rifat2022effect}. However, these parameters are more challenging to calculate which limits their widespread adoption. In the discussion, a more complex measure of surface roughness, entropy, is used to understand the mechanics of surface evolution during CDF.

\subsection{X-ray Characterization and Data Processing}

\begin{figure}[h]
      \centering \includegraphics[width=0.8\textwidth]{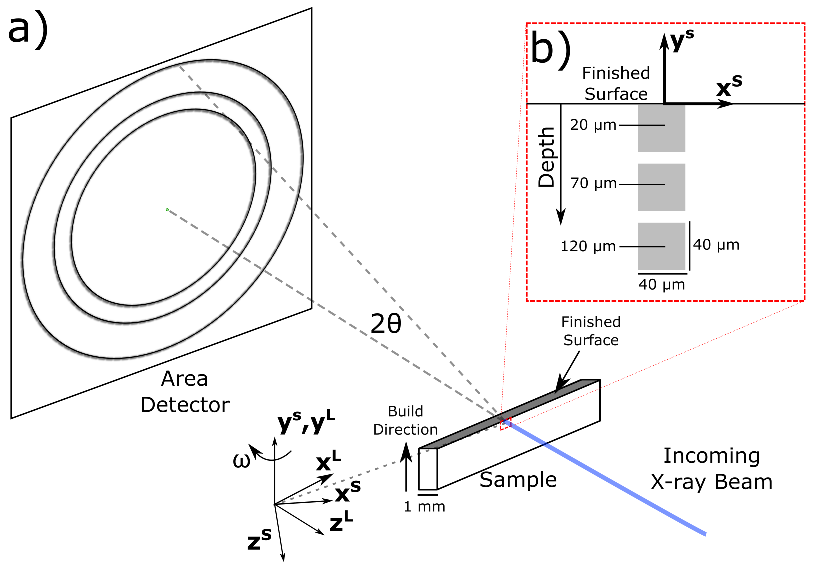}
    \caption{a) Experimental geometry for the X-ray diffraction measurements. The sample (S) and laboratory (L) coordinate systems are denoted. The sample is rotated by angle $\omega$ to probe various sample directions and measured diffracted intensity positions are quantified by the exit angle 2$\theta$. b) Inset showing the positions of diffraction volumes with respect to the finished surface in the probed specimens.}
    \label{fig:geom}
\end{figure}

Characterization of the subsurface state of the various specimens was performed at the QM$^2$ beamline of the Cornell High Energy Synchrotron Source \cite{nygren2020cartography}. A schematic of the measurements is provided in Fig. \ref{fig:geom}a. As the specimens were rotated during measurement, we define laboratory (superscript L) and sample (superscript S) coordinate systems that share a common vertical axis $\bm{y}^L$ and $\bm{y}^S$ that was also the specimen rotation axis. The current orientation of the sample coordinate system with respect to the laboratory is defined by the angle $\omega$. The specimen was placed for measurement such that build direction (BD) of the specimens were aligned with $\bm{y}^S$. The specimens were probed in a transmission geometry in which a 41 keV X-ray beam traveled in the $-\bm{z}^L$ direction. The beam size was 40 $\mu$m along $\bm{x}^L$ and 40 $\mu$m along $\bm{y}^L$.

Diffraction data from each specimen was collected at three depths with the X-ray beam centered at 20 $\mu$m, 70 $\mu$m, and 120 $\mu$m below the sample surface. A schematic of the X-ray measurement positions with respect to the sample surface are provided in Fig. \ref{fig:geom}b. Note that accounting for beam size, the top volume probes the sample surface. For measurements, diffraction images were collected continuously as the specimen was rotated by angle $\omega$ from -45$^\circ$ to 45$^\circ$ in 5$^\circ$ increments. The exposure time for each image (angular increment) was 5 s. The limited angular range was chosen to minimize the effects of absorption as the relatively planar samples were rotated and provided approximately 50\% pole figure coverage. Diffraction images were collected on a Pilatus 6M detector sitting 860 mm behind the specimen. The detector has 2527 pixel rows $\times$ 2463 pixel columns and a 172 $\mu$m pixel size. The detector was placed such that the 111, 200, and 220 diffraction rings were completely captured on the detector.

The diffraction images from the 5 specimens (3 volumes each) were processed using routines that take advantage of the HEXRD software package \cite{bernier2011far,nygren2020algorithm}. Each diffraction image was broken into 36 10$^\circ$ azimuthal bins and then each bin was azimuthally (around the ring) integrated to produce one-dimensional intensity line profiles. The azimuthal integration process consists of remapping each diffraction image to a polar coordinate system described by radial and azimuthal directions centered around the intersection of the incoming beam with the detector using bi-linear interpolation. The polar-remapped diffracted intensity is then integrated by summing intensity along the azimuthal direction. For each line profile, Pseudo-Voigt peaks \cite{young1993rietveld} (Gaussian and Lorentzian functions combined with a rule of mixtures) were fit to the 111, 200, and 222 diffraction peaks (sets of lattice planes). From the fits, integrated intensity and full-width-at-half-maximum (FWHM) were extracted. The intensity and FWHM from the diffraction peaks were then mapped to 111, 200, and 222 pole figures in the sample coordinate system. From the rotation step size and azimuthal bin size, each pole figure has 648 measurement points. The intensity pole figure data were then transferred to the MTEX software package \cite{hielscher2008novel} for calculation of orientation distribution functions (ODFs). The ODF calculated $f$ follows the standard definition \cite{kocks2000texture}
\begin{equation}
\frac{dV(\bm{r})}{V}=f(\bm{r})d\bm{r}
\end{equation}
where $\bm{r}$ is an orientation transformation from the local crystal to sample frame, $dV(\bm{r})$ is the volume fraction of crystal with orientation $\bm{r}$, and $V$ is the diffraction volume. ODFs are determined by minimizing the difference (L2 norm) between projections of the fit ODF onto synthetic pole figures and the measured pole figure data. Scalar crystallographic texture metrics are also calculated using MTEX including the strength $T$ of the crystallographic texture defined as
\begin{equation}
T=\int f(\bm{r})^2d\bm{r}
\end{equation}
and the entropy $S$, to quantify the disorder, defined as 
\begin{equation}
S=-\int f(\bm{r}) \ln{(f(\bm{r}))} d\bm{r} \quad .
\end{equation}
Lastly, the fit ODFs were re-projected to pole figures for crystallographic texture data presentation.

\section{Results}

\subsection{Surface Measurements}
\label{surface_measurements}
\begin{figure}[h]
      \centering \includegraphics[width=1.0\textwidth]{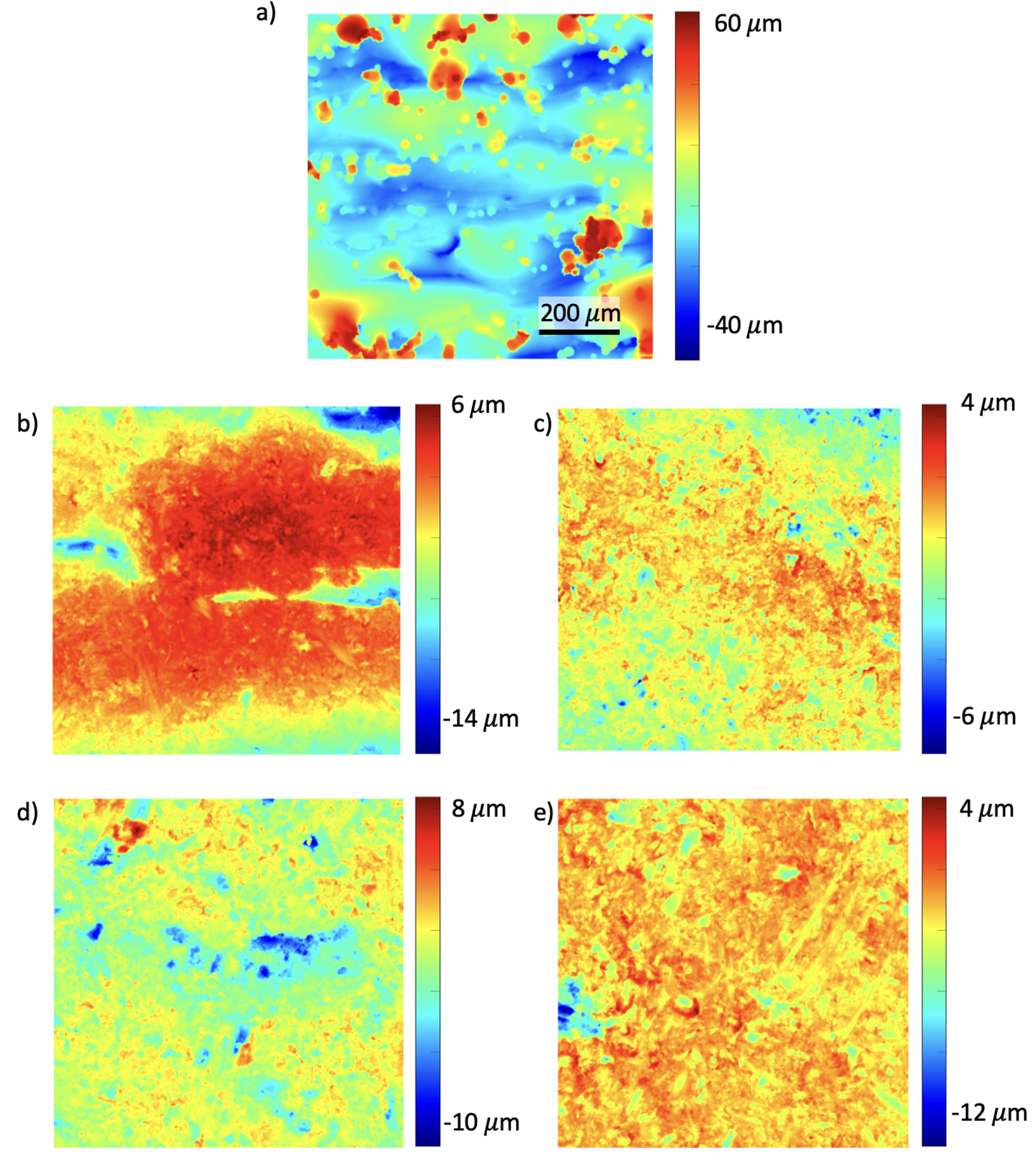}
    \caption{Surface topographies of representative specimens in the following stages of CDF: a) NF (not finished), b) U1 (6$\times$6 SCT media, 75 mins), c) U2 (6$\times$6 SCT media, 135 mins), d) B1 (bimodal media, 75 mins), e) B2 (bimodal media, 135 mins).}
    \label{surface_topographies}
\end{figure}

The surface in the NF state exhibited a rough texture that featured roughness measures of $R_a\sim4.9\ \mu$m, $R_q\sim 6.5\ \mu$m, and $R_z\sim52.7\ \mu$m. An example of the texture resulting from AM is shown in Fig. \ref{surface_topographies}a. As expected, this NF texture contains waviness that often results from the process of additive (layer-by-layer) fabrication. In addition, the figure also shows randomly positioned peaks each spanning a much smaller length scale of approximately 20 to 80 $\mu$m diameter. These peaks originated from partially melted powder particles that adhere to the surface during AM. 

In comparison to the NF specimen, the surfaces that underwent CDF showed considerable reduction in surface roughness measures. Example surface profiles of the finished specimens are given in Figs. \ref{surface_topographies}b-e, while a summary of these measures is given in Table \ref{tab:roughness_data}. After 75 mins of processing with the unimodal WXC 6$\times$6 media (U1), the surfaces exhibited $R_a=2.27\ \mu$m, $R_q=2.78\ \mu$m, and $R_z=18.74\ \mu$m. After 60 more mins of processing for a total of 135 mins (U2), these measures further reduced significantly and exhibited $R_a=0.79\ \mu$m, $R_q=1.06\ \mu$m, and $R_z=17.47\ \mu$m. 

The bimodal media mixture that comprised a 50:50 mixture of WXC 6$\times$6 SCT and WXC 15$\times$15 ACT media (B1) showed more rapid improvements compared with the surfaces processed using unimodal media after 75 mins of processing. The surfaces exhibited $R_a=1.15\ \mu$m, $R_q=1.48\ \mu$m, and $Rz=14.83\ \mu$m. After 135 mins of processing (B2), the arithmetic and root mean square measures of surface roughness further reduced to $R_a=0.97\ \mu$m, $R_q=1.36\ \mu$m. However, now the maximum peak to valley height exhibited a slight increase to $Rz=20.75\ \mu$m.

The repeatability of the measurements obtained using white light profilometry is provided in the error bars (2 $\times$ the error $e$) associated to the quantified surface roughness metrics ($R+e$) which are reported in Table \ref{tab:roughness_data}. These error bars were characterized from 3 different measurements that were made on 3 different randomly chosen points on every specimen under equivalent conditions. A maximum error of $2e = 1.30$ $\mu$m was determined for surface roughness metric $R_a$ that was obtained from the U2 specimen prior to finishing. We note that this value is significantly larger than the resolution of the optical profilometer (reported as several angstroms) and is likely a consequence of how rough the surfaces themselves were in this state.

We anticipate that the error associated with surface roughness metrics would inflate if sufficiently large areas were not measured during white light profilometry. Nonetheless, increasing the dimensions of the measured regions also comes at the cost of decreasing spatial resolution of the measured surface topographies. To this end, we note that the dimensions of our measured regions ($\sim$ 0.85 mm $\times$  0.85 mm) were similar to those in other well-established works within this focus area ($\sim$ 1.5 mm $\times$ 1 mm) \cite{kahlin2020improved}.

\begin{table}[h]
\footnotesize
\centering

\caption{Roughness characteristics $R_a$, $R_q$ and $R_z$ of the finished AM Inconel 718 samples including pre- and post- finishing (initial and final respectively).}

    \begin{tabular}{|c|c|c|c|c|c|c|}
         \hline
         \textbf{Name} &  \multicolumn{2}{c|}{$R_a$} &  \multicolumn{2}{c|}{$R_q$}  &  \multicolumn{2}{c|} {$R_z$} \\
         \hline
         & Initial & Final & Initial & Final  & Initial & Final \\
         \hline
         U1&4.93$\pm$0.23 & 2.27$\pm$0.54 & 6.44$\pm$0.13 & 2.78$\pm$0.61 & 56.13$\pm$6.69& 18.74$\pm$3.04  \\
         U2&5.25$\pm$0.65 & 0.79$\pm$0.15 & 6.85$\pm$0.89 & 1.06$\pm$0.17 & 52.06$\pm$2.81& 17.47$\pm$5.90 \\
         \hline
         B1&4.39$\pm$0.33 & 1.15$\pm$0.24 & 6.04$\pm$0.44 & 1.48$\pm$0.28 & 50.54$\pm$1.96& 14.83$\pm$2.33  \\
         B2& 4.82$\pm$0.13 & 0.97$\pm$0.05 & 6.53$\pm$0.17 & 1.36$\pm$0.12 & 54.94$\pm$3.23& 20.75$\pm$6.92  \\
         \hline
    \end{tabular}

\label{tab:roughness_data}
\end{table}


\subsection{X-ray Measurements}

\begin{figure}[h]
      \centering \includegraphics[width=0.75\textwidth]{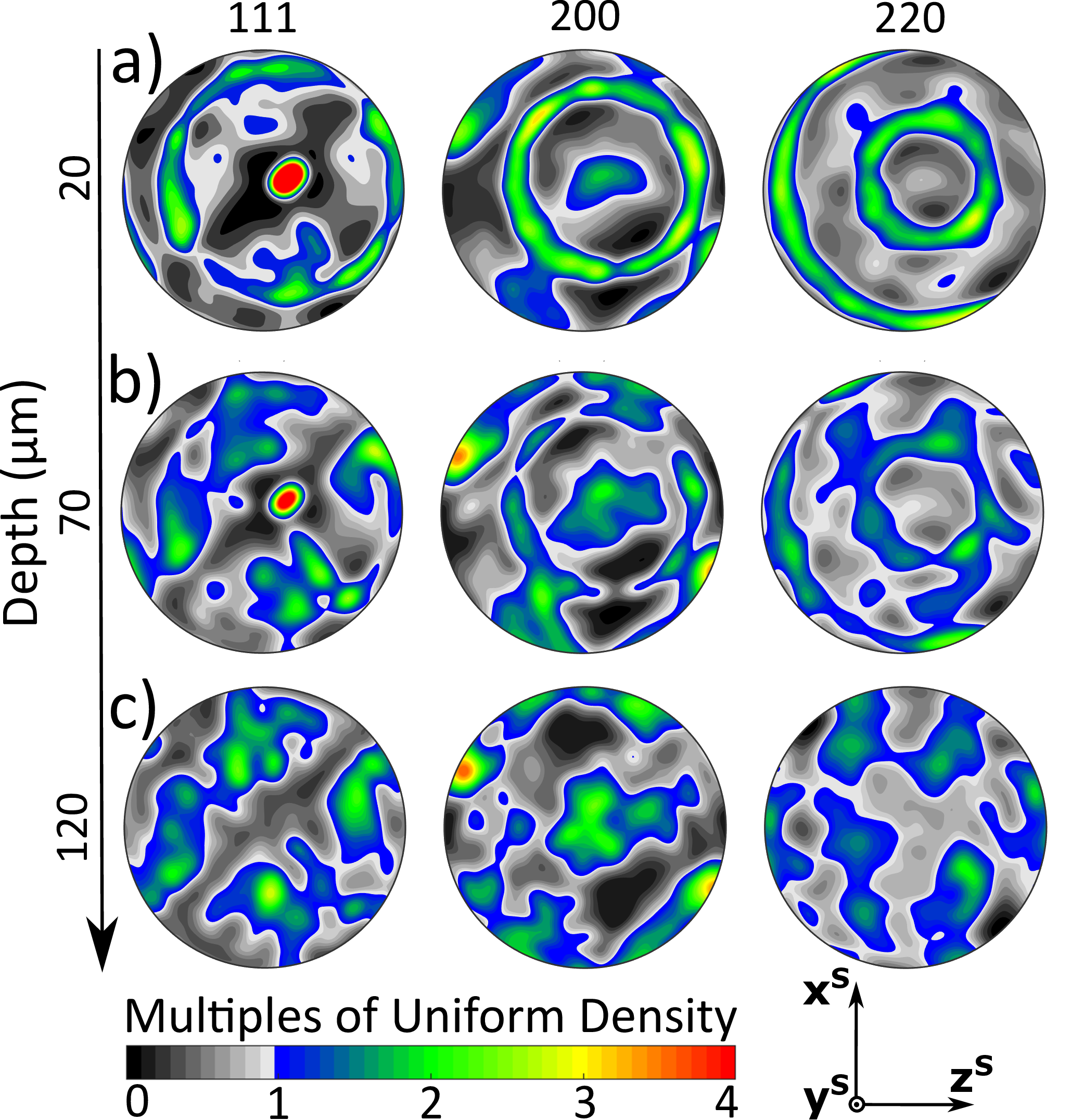}
    \caption{Comparisons of the measured crystallographic textures in the NF (no finish) specimen at various depths: a) 20 $\mu$m, b) 70 $\mu$m, and c) 120 $\mu$m. }
    \label{fig:nf1_texture}
\end{figure}

As described, the crystallographic textures of the samples were measured at three depths away from the surface. Pole figures from the 111, 200, and 220 lattice planes in the NF specimen are projected from the ODFs calculated and presented in Fig. \ref{fig:nf1_texture} as a function of depth: 20 $\mu$m (Fig. \ref{fig:nf1_texture}a), 70 $\mu$m (Fig. \ref{fig:nf1_texture}b), and 120 $\mu$m (Fig. \ref{fig:nf1_texture}c). The orientation pole figure data is presented in multiples of uniform density (MUD).  In Fig. \ref{fig:nf1_texture}, we see that the NF specimen has a relatively strong fiber texture at the sample surface (20 $\mu$m depth) with the \{111\} lattice plane normals nominally aligned with the build direction and the sample surface. The grains in this diffraction volume are randomly oriented about the [111] direction as evidenced by the rings of intensity observed in the 200 and 220 pole figures. With increasing depth, the texture transitions from the 111 fiber texture to a weak cube texture which we can see from the larger MUD values on the 200 pole figures along the $\bm{y^S}$ (build) direction and the 4-fold symmetry of intensity about $\bm{y^S}$ on all three pole figures.

\begin{figure}[h]
      \centering \includegraphics[width=0.75\textwidth]{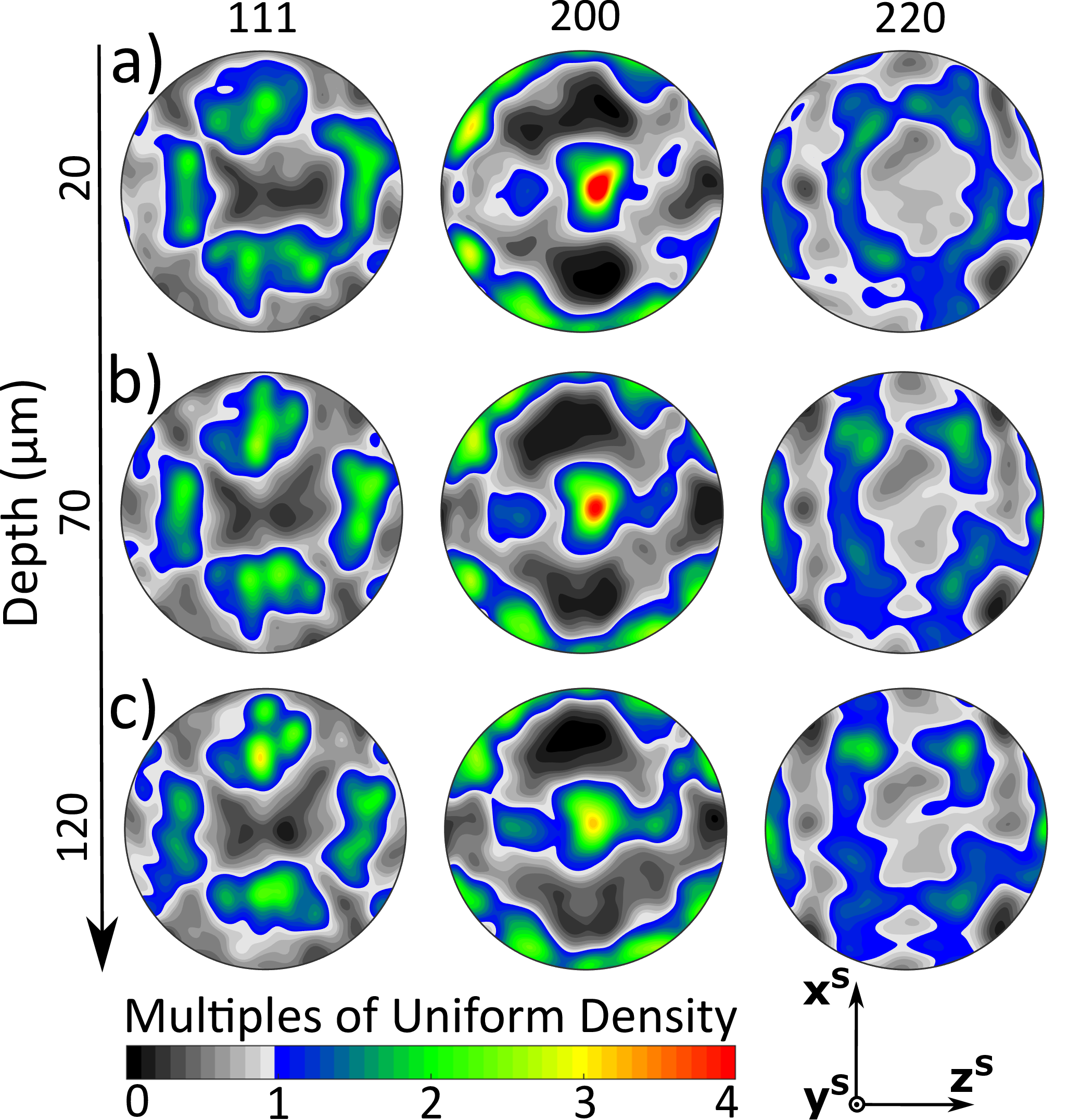}
    \caption{Comparisons of the measured crystallographic textures in the U1 (6$\times$6 SCT media, 75 mins) surface-finished specimen at various depths: a) 20 $\mu$m, b) 70 $\mu$m, and c) 120 $\mu$m. }
    \label{fig:u1_texture}
\end{figure}

\begin{figure}[h]
      \centering \includegraphics[width=0.75\textwidth]{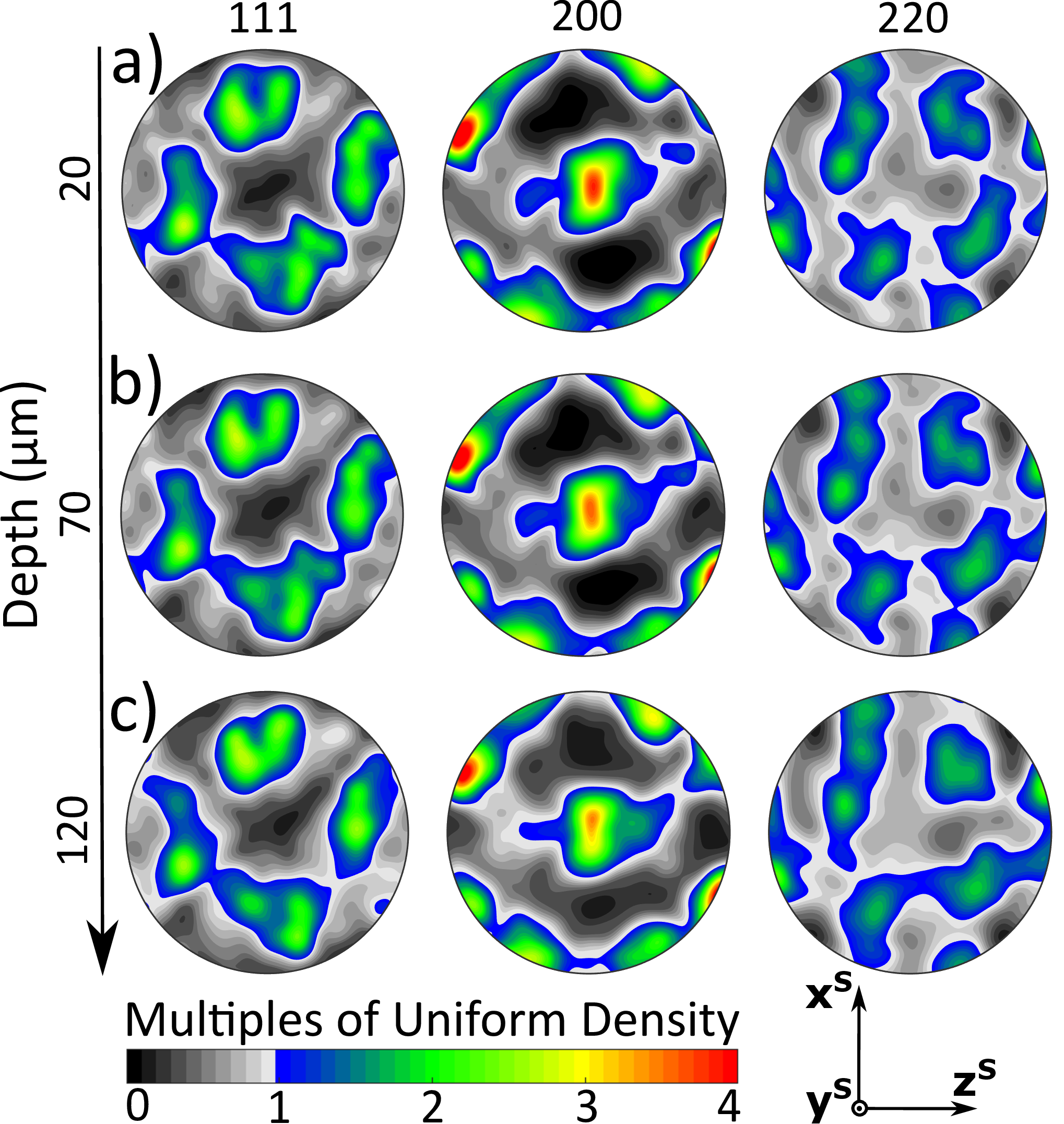}
    \caption{Comparisons of the measured crystallographic textures in the B1 (bimodal media, 75 mins) surface-finished specimen at various depths: a) 20 $\mu$m, b) 70 $\mu$m, and c) 120 $\mu$m. }
    \label{fig:b1_texture}
\end{figure}

Similar to the NF sample, texture data for the U1 and B1 specimens (see Table \ref{tab:rough_table}) are shown in \ref{fig:u1_texture} and Fig. \ref{fig:b1_texture} respectively. The textures for the U2 and B2 specimens are not presented as they are qualitatively similar to U1 and B1. All measured textures with depth in both  U1 and B1 specimens (Fig. \ref{fig:u1_texture} and Fig. \ref{fig:b1_texture} respectively) are relatively similar. Each exhibit a cube texture similar to the lowest depth (120 $\mu$m) in Fig. \ref{fig:nf1_texture}c, but with higher peak values. In both samples, the peak intensity values in the pole figures appear to decrease with decreasing depth, implying the surface treatments reinforce the existing cube texture from the AM build. The lack of the 111 fiber texture in the CDF-finished samples indicates that approximately 100 $\mu$m of material was removed from the samples and along with it, the remnant surface texture from the AM build process.

\begin{figure}[h]
      \centering \includegraphics[width=1.0\textwidth]{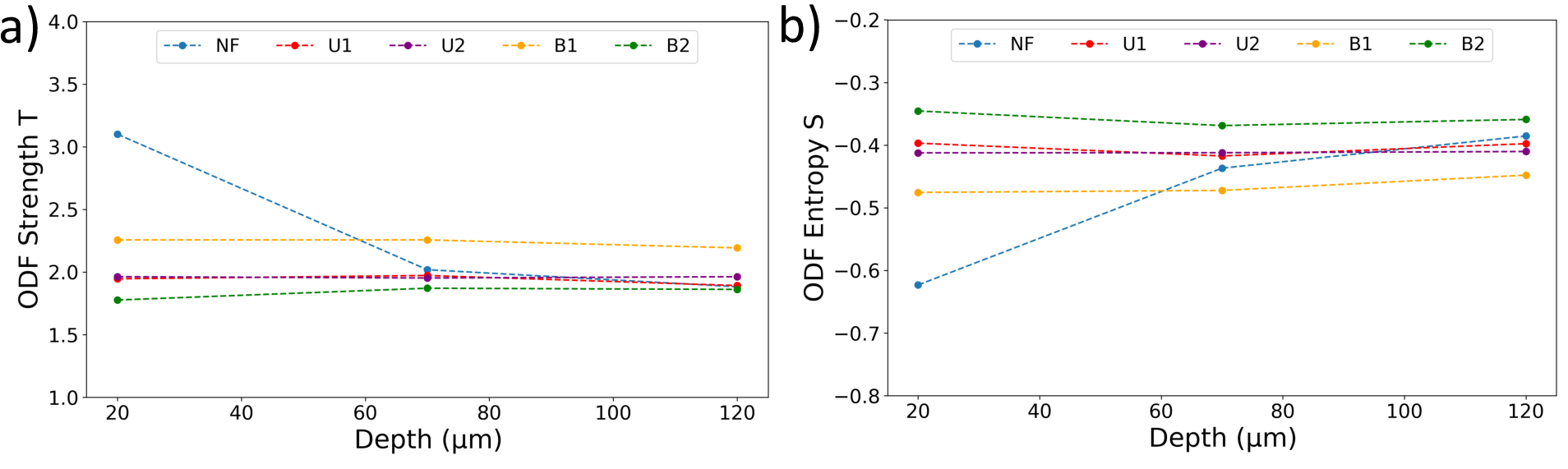}
    \caption{a) Variation of the ODF strength $T$ in the five specimens as a function of depth. b) Variation of the ODF entropy $S$ as a function of depth.}
    \label{fig:texture_depth}
\end{figure}

Figure \ref{fig:texture_depth}a shows the variation of the ODF strength $T$ with depth. We can see that in the NF specimen, the strongest crystallographic texture is at the sample surface which decreases with depth, eventually reaching crystallographic texture strengths comparable to the surface-finished specimens. This is consistent with the abrasive media removing the top surface layer of the AM build. In the surface finished specimens, the crystallographic texture strengths are comparable, with the B1 texture being the strongest. However, the relatively constant crystallographic texture strength with depth implies that the increased strength B1 is most likely a remnant of the initial crystallographic texture of the AM build. The crystallographic texture strengths appear to peak at the 70 $\mu$m depth, in all specimens but this increase is small and within measurement uncertainty. Similar findings, but reversed, can be seen in Fig. \ref{fig:texture_depth}b showing the variation ODF entropy $S$ with depth.

To examine the microscale residual strains present in the material, the full-width-at-half-maximum (FWHM) of the various sets of lattice planes with respect to sample direction were also extracted from the diffraction data. The FWHM of a diffraction peak captures the spread of lattice plane spacing across a diffraction volume which will generally have contributions from stress features across length scales. In this case, the peak FWHM primarily has contributions from residual stresses generated during grain-scale plastic deformation and local defect (dislocation) content. We note that as the samples were crystallographic textured, FWHM values from diffraction peaks with negligible intensity are omitted.

\begin{figure}[h]
      \centering \includegraphics[width=1.0\textwidth]{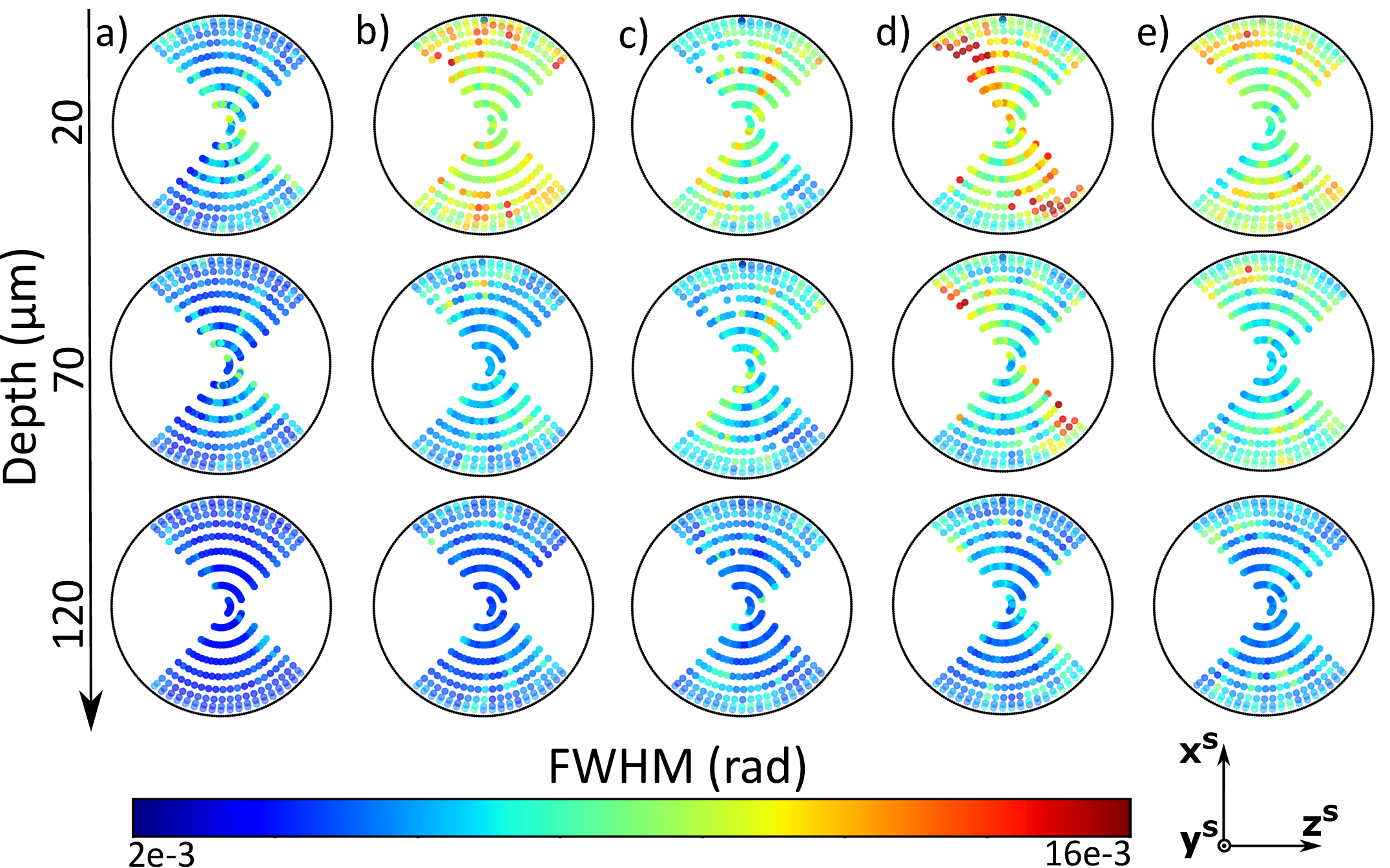}
    \caption{Comparison of 200 diffraction peak FWHM as a function of sample direction and depth below the finished surface. Data is presented from the a) NF (not finished), b)  U1 (6$\times$6 SCT media, 75 mins), c) U2 (6$\times$6 SCT media, 135 mins), d) B1 (bimodal media, 75 mins), e) B2 (bimodal media, 135 mins) specimens.}
    \label{fig:fwhm_200}
\end{figure}

As an example of the FWHM data, Figure \ref{fig:fwhm_200} shows the variation of anisotropic FWHM of the 200 diffraction peak as a function of the depth for the a) NF, b) U1, c) U2, d) B1, and e) B2 specimens. Similar distribution trends are observed in the 111 and 220 pole figures (not shown), but with different magnitudes associated with increases in FWHM as a function of 2$\theta$ and dislocation contrast factors \cite{wilkens1970determination}. Each point on these 200 pole figures correspond to the spread of the diffraction peaks in the 2$\theta$ direction on the detector (see Fig. \ref{fig:geom}a) mapped to a sample direction. A larger value indicates that lattice planes with normal parallel to the sample direction have a larger spread of spacing (and elastic strain). The gaps correspond to the -45$^\circ$ to 45$^\circ$ angular range across which the sample was scanned. In general, the FWHMs and associated residual microscale strains are much larger in the finished specimens. There is also a decrease in FWHM with depth observed in all specimens. Regarding anisotropy, the FWHM values are generally largest perpendicular to $\bm{y^S}$ (surface / build direction) which correspond to the points around the equator of the pole figure.The converse is true for the NF specimen, the largest FWHM values are parallel to $\bm{y^S}$ which are the points at the center of the pole figure.

\begin{figure}[h]
      \centering \includegraphics[width=1.0\textwidth]{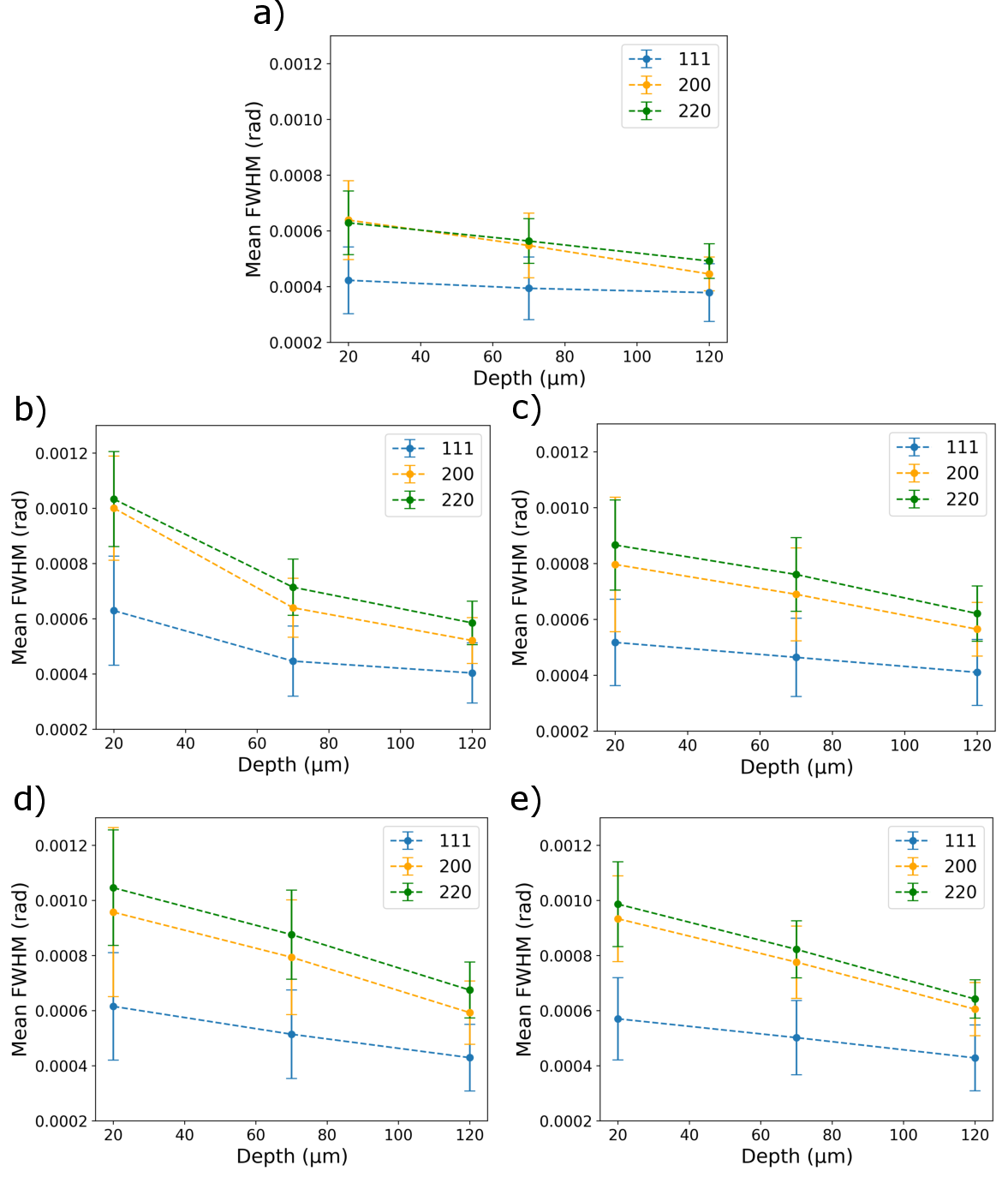}
    \caption{Comparison of the mean (data points) and standard deviation (error bars) of anisotropic diffraction peak FWHM with depth below the finished surface. Data from the 111, 200, and 220 diffraction peaks are shown. Data is presented from the a) NF (not finished), b)  U1 (6$\times$6 SCT media, 75 mins), c) U2 (6$\times$6 SCT media, 135 mins), d) B1 (bimodal media, 75 mins), e) B2 (bimodal media, 135 mins) specimens.}
    \label{fig:fwhm_depth}
\end{figure}

To examine these trends more quantitatively, Fig. \ref{fig:fwhm_depth} shows the variation of mean (points) and standard deviation (error bars) of FWHM for the 111, 200, and 220 across sample directions. The subfigures are associated with the a) NF, b) U1, c) U2, d) B1, and e) B2 specimens. In all specimens, besides U1, the decreases in FWHM with depth are fairly linear. The FWHM values (and underlying strain) are about 50\% higher in the finished specimens in comparison to the NF specimen. At 120 $\mu$m, the finished specimens also begin to approach the same magnitudes of FWHM as the NF, albeit still slightly higher. Importantly, all of the mean FWHM magnitudes in the finished specimens are similar, indicating the exact nature of the media and finishing time do not generate appreciable differences in surface residual strain (and stress) states, at least for the conditions examined.

\section{Discussion}

Microstructure at the surface will naturally play a critical role in governing the functional response of a part. An example of such surface microstructure-dependent functional response is the ability to withstand damage during fatigue loading \cite{mcdowell2010microstructure}. Unfortunately, such functional response is complicated by the influence of roughness textures on the surface, via stress-concentrations that originate at the valleys between roughness peaks. In this regard, classical destructive methods of sample preparation such as serial-sectioning are inadequate to simultaneously explore the development of surface strains and roughness textures during complex finishing processes. Hence, high-energy X-ray diffraction was used to characterize the variation of microstructure with distance from the surface in AM Inconel 718, with and without surface finishing. These measurements provide a unique means to non-destructively probe near the sample surface with the benefit of reducing the amount of alteration to the surface residual strain state in comparison to serial section methods.

Perhaps the most surprising finding in this work was the similarity in microstructures and residual strain across finishing conditions. The levels of residual strain imparted into the specimen at the surface are fairly comparable  (an approximately 50\% increase from the specimen with no finish) across the media and finishing times. This indicates that while changes to the microstructure at the surface need to be considered (particularly removal of any surface microstructures resulting from AM), there may be limited opportunities for microstructural optimization through CDF. This finding can also be stated in the inverse, focus can be placed on optimizing surface finish (media selection and time) without fear of inadvertently creating a particularly deleterious microstructure. However, we acknowledge that further work is required to determine the generality of the findings. In particular, here we varied the media and finishing time, but the disk speed can also be varied. A change in disk speed will alter both the flow path of the media and parts, as well as change their relative velocity and contact pressure, which could alter underlying microstructure.

The finishing mechanism of CDF involves a combination of shear-based material removal and smearing of surface roughness asperities through plastic deformation, both of which result in smaller measures of surface roughness. These mechanisms involve unique trajectories of microstructure evolution that produce different microstructural consequences. Furthermore, the exact mechanism is dynamically governed by the local mechanical conditions during CDF. Here the results indicate that the amount of plastic deformation induced by CDF is moderate (at least for the processing conditions examined). The increase in diffraction peak widths observed near the surface is tied to an increase in local dislocation content and deformation compatibility which occurs during plastic deformation. However, the amount of plastic deformation was not sufficient to induce large-scale grain rotation or dynamic recrystallization which would lead to the appearance of new crystallographic texture components in the pole figures in comparison to the not-finished condition. These observations are in contrast to most shear-based material removal processes which are usually classified as surface severe plastic deformation processes (like machining and burnishing). Recent efforts have attempted to phenomenologically model these phenomena by investigating the coupled evolution of part surface roughness and the media condition with respect to CDF processing time \cite{rifat2024evolution}. As the removed surface material here had a relatively strong fiber texture, which will naturally have a strongly anisotropic mechanical response, the effects of its removal will be load-condition dependent. The increased defect content and accompanying work-hardening at the surface will produce a moderate `case-hardening' effect that can be beneficial in high-cycle fatigue applications.

The results described in section \ref{surface_measurements} suggest avenues for process optimization for achieving an ideal surface texture within a short duration. The 50:50 bimodal mixture of media was able to provide an appreciable improvement in measures of surface roughness, e.g., $R_a=$ 1.15 $\mu$m after 75 mins of processing compared with the starting value of $R_a=$ 4.9 $\mu$m before finishing. This measure would be eventually superseded by the smaller 6$\times$6 SCT media that provided $R_a=$ 0.79 $\mu$m after 135 mins of finishing. Utilizing the bimodal media alone would have provided $R_a=$ 0.97 $\mu$m after 135 mins of processing. This suggests that dynamically altering the media from bimodal to unimodal may facilitate an adequate amount of processing in a shorter amount of time. Nonetheless, such studies do not provide a comparison of the surface resulting from such processing routes beyond a trivial comparison of its surface roughness measures like $R_a$, $R_q$, and $R_z$. This detail has been realized to be critical for formulating predictive models of part functional response \cite{rifat2020microstructure,rifat2022effect}. To bridge this knowledge gap,  a detailed analysis of the surfaces resulting after the equivalent processing times but with different media is attempted in the next section.

\subsection{Fourier Analysis}
To further characterize the salient features of the surface resulting from CDF, their spectral entropies were quantified. This parameter provides a measure of the disorder of the surface topography. Order in a surface can be envisioned as the regularity of its features. For instance, a flat and perfectly smooth surface is perfectly ordered and features an entropy of 0. The same is true for a pure sinusoidal surface. However, surfaces with complex topographies are much more disordered and feature higher ($\leq$ 1) entropies. The disorder in the topography of a surface is a consequence of the manufacturing process route with which it was fabricated. In this regard, the spectral entropies of surfaces resulting from CDF can provide a measure of the disorder in the mechanics of interaction between media (uni/bi modal) and the specimen surface. 

The spectral entropy $H$ of a 1D spectrum is given by
\begin{equation}
H=\frac{-\sum_{k=1}^{N} p_k\log_2 p_k}{\log_2 N}
\label{spectral_entropy}
\end{equation}
where
\begin{equation}
p_k=\frac{||A_k||^2}{\sum_{k=1}^N ||A_k||^2} \quad ,
\end{equation}
$A_k$ is the amplitude of the Fourier series component of wavenumber $k$, and $N$ is the length of the signal. This formulation was adapted in this work to calculate the spectral entropy of a 2D spectrum. This was done by vectorizing the 2D spectrum obtained by fast Fourier transform on the surface from a 2D array to 1D list of amplitudes, and subsequently using the formulation described in this Eqn. \ref{spectral_entropy}. Spectral entropies were calculated from the white light profilometry data from which the measures of surface roughness were also calculated. Scattered missing pixels in the profilometry measurements were filled by averaging the neighboring points using the `griddata' function in Matlab.

\begin{figure}[h]
      \centering \includegraphics[width=1.0\textwidth]{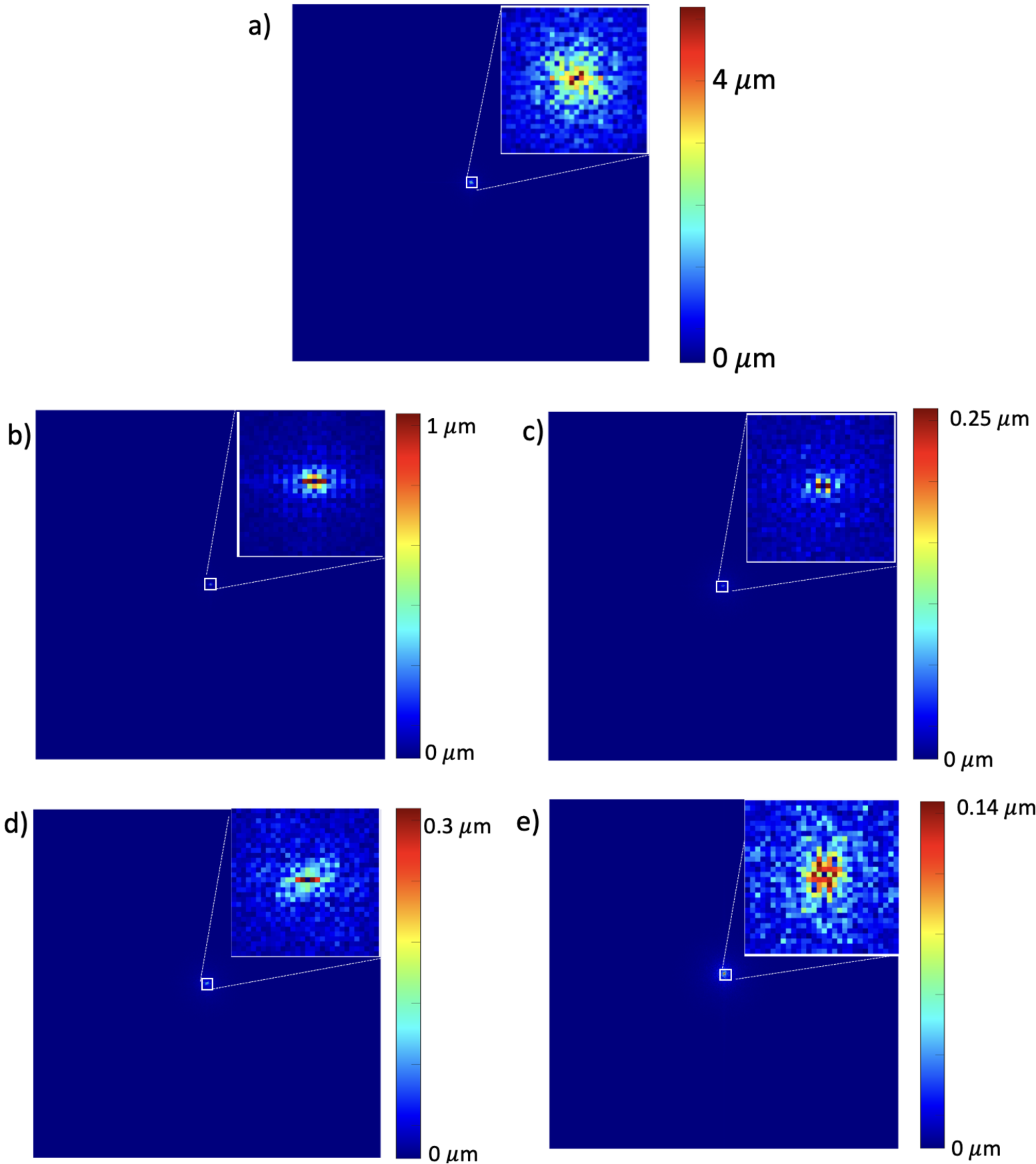}
    \caption{Fourier series analysis of surface topographies of representative specimens in various stages of CDF. a) NF (not finished), b)  U1 (6$\times$6 SCT media, 75 mins), c) U2 (6$\times$6 SCT media, 135 mins), d) B1 (bimodal media, 75 mins), e) B2 (bimodal media, 135 mins). Spectrums are center-shifted and correspond to the original surface textures shown in Fig. \ref{surface_topographies}. The insets show closeups of the central portions of these spectrums. Dimensions of the insets are 31 $\times$ 31 wavenumbers.}
    \label{fourier_analysis}
\end{figure}

Figure \ref{fourier_analysis} shows representative spectrums of the NF (Fig. \ref{fourier_analysis}a ), U1 (Fig. \ref{fourier_analysis}b), U2 (Fig. \ref{fourier_analysis}c), B1 (Fig. \ref{fourier_analysis}d), and B2 (Fig. \ref{fourier_analysis}e) surfaces with insets showing the low wavenumber regions in more detail. In the figure, we see the NF specimen surface has a high amount of disorder, evidenced from the high amplitudes of waves with low wavenumbers (inset in Fig. \ref{fourier_analysis}a). This disorder originates from the layer-by-layer nature of additive manufacturing which produces semi-regular undulations on the surface. In addition, partially melted powders also fuse to the surface during AM, which can be an additional source of the disorder in this surface. The quantification of spectral entropy of this surface suggested $H=0.43$. Upon finishing with the 6$\times$6 SCT media for 75 mins, (U1), the specimen surface assumed a less disordered state evidenced in its much lower spectral entropy of H=0.27 (Fig. \ref{fourier_analysis}b). However, upon further finishing with the same media to a total of 135 mins, the spectral entropy of the surface increased to H=0.47 (U2). In comparison, finishing with the bimodal media resulted in increase in spectral entropy to H = 0.48 (B1, 75 mins), and H = 0.52 (B2, 135 mins). 

The results summarized in the previous paragraph suggest that the surfaces resulting from CDF with bimodal media were always more disordered than surfaces resulting from finishing with unimodal media. This disorder can originate from multiple sources. The first phase of finishing by the 6$\times$6 media for 75 mins, (U1) reduces the disorder compared with the NF surface, in-line with the also observed reduction in surface roughness measures. However, the additional 60 mins of finishing with the same 6$\times$6 media (U2) results in a further decrease in surface roughness and an increase in the spectral entropy. This implies that the mechanics of surface smoothing underwent a transition between 75, and 135 mins, which resulted in this increase of disorder. This transition was accompanied by an increase in the magnitude of residual stresses at $\sim$70 $\mu$m for the U2 specimen. A similar transition was however not seen in the bimodal media that exhibited a decrease in surface roughness and increase in spectral entropy throughout the finishing process.

The transition may arise from evolution of media characteristics \cite{rifat2024evolution} during the finishing process, which occurs due to self interactions. Such self interactions are also likely in the bimodal media. However, their effect is further complicated by the high momentum of the 15$\times$15 media which can also produce a greater surface smoothing effect and residual stresses.

\section{Summary and Conclusions}

A unique combination of traditional surface profilometry and synchrotron high-energy X-ray diffraction were combined to build a more complete picture of surface roughness reductions and changes to microstructure below the surface in AM Inconel 718. Four different centrifugal disk finishing conditions (two media types $\times$ two finishing times) were examined along with a reference specimen. From the characterization, we found:
\begin{enumerate}
    \item While bimodal abrasive media was more effective at improving surface finish, both the unimodal and bimodal abrasive media altered the underlying microstructure and residual microscale strain in a similar fashion.
    \item Both abrasive media types removed strongly (fiber) textured material at an AM surface, leaving only bulk (cube) textures without the introduction of any new crystallographic texture components.
    \item The abrasive finishing imparts a significant amount of residual microscale strains at a sample surface which decays within 100 $\mu$m. 
    \item The unimodal media produces first a decrease in disorder of the surface topography, and then an increase in this disorder. In comparison the bimodal media always provides a higher level of disorder in the surface. These trends match those observed in surface roughness measures, and residual stresses. 
\end{enumerate}
The results here point to the need for future work to understand the effect of preexisting geometries on the mechanics of CDF and their effects of fatigue life. Complicated geometries will significantly influence the mechanics of flow of the media during CDF, and thereby alter its process consequences. Without a thorough understanding of these underpinnings, it is not clear how process optimization can be achieved for parts with complex geometries, such as those commonly originating from additive manufacturing. In addition, CDF appears to provide an interesting combination of material removal with only moderate plastic deformation. How this combination affects fatigue life should be further explored.

\section{Acknowledgements}
Dr. Jacob Ruff is thanked for his help performing the diffraction measurements. This material is based upon work supported by the National Science Foundation under Grant No. 1825686. This work is based on research conducted at the Center for High-Energy X-ray Sciences (CHEXS), which is supported by the National Science Foundation (BIO, ENG and MPS Directorates) under award DMR-1829070.

\bibliography{bibliography}
\end{document}